\documentclass[epj,final]{svjour}

\usepackage{latexsym}
\usepackage{url}
\usepackage{amsfonts}
\usepackage{tikz}
\usetikzlibrary{decorations.pathmorphing}
\usetikzlibrary{decorations.pathmorphing,calc}
\usepackage{amsmath, amssymb}
\RequirePackage{graphicx}

\begin{document}
\title{Cosmological constant, information and gedanken experiments with black hole horizons}
\author{V.G. Gurzadyan\inst{1,2}, A.~Stepanian\inst{1}
}                     
%
%
\institute{Center for Cosmology and Astrophysics, Alikhanian National Laboratory and Yerevan State University, Yerevan, Armenia \and
SIA, Sapienza Universita di Roma, Rome, Italy}
\date{Received: date / Revised version: date}
%

\abstract{The cosmological constant if considered as a fundamental constant, provides an information treatment for gravitation problems, both cosmological and of black holes. The efficiency of that approach is shown via gedanken experiments for the information behavior of the horizons for Schwarzschild-de Sitter and Kerr-de Sitter metrics.  A notion of entropy regarding any observer and in all possible non-extreme black hole solutions is suggested, linked also to Bekenstein bound. The suggested information approach forbids the existence of naked singularities.}

\PACS{
      {98.80.-k}{Cosmology}   
     } 
%
\maketitle

\section{Introduction}

Einstein denoted the cosmological constant $\Lambda$ as a "universal constant" in his original papers  on the cosmological model \cite{E,E1}. The recent observational indications on the accelerated expansion of the Universe essentially revived the multi-facet problem of the cosmological constant. Here we will deal with the consequences of the consideration of cosmological constant in the context of other physical constants \cite{Uz}.

In \cite{GS1} it was shown that the cosmological constant can be considered as a second gravitational constant, along with the Newtonian constant, as it is dimension-independent and matter uncoupled. That description follows from the weak-field General Relativity involving the cosmological constant \cite{GS1,G1}. The latter is based on the general function satisfying the Newton's theorem on the equivalency of the gravity of sphere and point mass \cite{G}. Then, the metric involves both constants, $G$ and $\Lambda$, and is defined as follows
\begin{equation}\label{mod}
g_{00} = 1- \frac{2GM}{c^2 r} - \frac{\Lambda r^2}{3}; \quad g_{rr} = (1- \frac{2GM}{c^2 r} -\frac{\Lambda r^2}{3})^{-1},
\end{equation}
i.e. in the form of Schwarzschild-de Sitter spherically-symmetric metric. As weak-field General Relativity it was applied to the dynamics of groups and clusters of galaxies \cite{GS1,G1,GS2}, thus enabling one to describe the observational data on the dark matter in those cosmic structures.

The consideration of the cosmological constant $\Lambda$ as the fourth fundamental constant - along with $c$, $\hbar$ and $G$ - opens new principal possibilities to treat the cosmological evolution and the black hole event horizons \cite{GS4}. 

Namely, the combinations of four fundamental constants leads to the appearance of dimensionless quantities \cite{GS4} 
\begin{equation}\label{Nat}
m I^n= m \frac{c^{3n}}{\hbar^n G^n \Lambda^n}, \quad m, n \in \mathbb{R}.
\end{equation}
For $n=1$ and $m=3\pi$ this relation represents the notion of information  - or of entropy with Boltzmann constant $k_B$ -  of de Sitter (dS) universe \cite{HG,BekB}. 

According to Bekenstein \cite{Bek} the ``simplest" particle in the Universe has an area equal to $\frac{4 G \hbar}{c^3}$ which contains exactly 1 unit  of information. That notion can be generalized to cosmological horizons too \cite{HG}. Thus, by considering the Hubble horizon $\frac{c}{H}$, where $H$ is the Hubble constant obtained from Friedmann equations
\begin{equation}\label{FLRW}
H(t)^2 = \frac{8 \pi G \rho}{3} + \frac{\Lambda c^2}{3} -\frac{k c^2}{a(t)^2},
\end{equation}
one can conclude that once the total area of the Hubble horizon was $4 l_p^2$, its Information Content (IC) equals exactly 1. In addition, it can be checked that inside the Hubble horizon the Bekenstein bound
\begin{equation}\label{BBdS}
I_{BB} \leq  \frac {2 \pi R E}{ c \hbar ln 2} \quad bits, \quad S_{BB}\leq  \frac {2 \pi R E k_B}{ c \hbar},
\end{equation}
where $R$ and $E$ represent the characteristic radius and energy of the universe, respectively, is always saturated. Consequently, by considering Eqs.(\ref{FLRW}, \ref{BBdS}), one can conclude that once the total density of the Universe inside the Hubble horizon becomes $\rho_{tot} = \frac{3 c^5}{8 i \hbar G^2}$, the IC is increased by one unit i.e.
\begin{equation}\label{BBI}
I_{BB}[\frac{3 c^5}{i 8 \hbar G^2}]= i, \quad i=1,...,3 \pi I.
\end{equation}
In this sense, the evolution of Universe can be regarded as a discrete process, when the first unit of information is created when the area of Hubble horizon was $4l_p^2$ and its density was $\frac{3}{8} \frac{c^5}{\hbar G^2}$, accordingly. Consequently, at the $i$-th step, where the area is grown to $4 i l_p^2$ and the density becomes $\frac{3}{8 i} \frac{c^5}{\hbar G^2}$, the IC of universe becomes $i$. The process continues until the area of Hubble horizon reaches $\frac{12 \pi}{\Lambda}$ i.e. the  dS cosmological horizon and the density become equal to $\frac{\Lambda c^2}{8 \pi G}$, accordingly. 

In this paper, considering $\Lambda$ as a fundamental constant, we investigate the information aspects of the black hole event horizons. Within that information treatment we use gedanken experiments to deal with the Schwarzschild-de Sitter and Kerr-de Sitter black holes and Hawking effect. The notion of entropy follows that approach and its link to Bekenstein bound forbid the possibility of existence of naked singularities.

\section{Horizons for Schwarzschild-de Sitter and Kerr-de Sitter black holes}

Considering $\Lambda$ as one of the fundamental constants, it is clear that it should enter in the BH solutions. Particularly, the two astrophysical candidates of BHs i.e. Schwarzschild and Kerr solutions will be modified accordingly. Namely, the metric of Schwarzschild BH changes to the Schwarzschild-de Sitter (SdS)
\begin{equation}\label{SdS}
ds^2 = (1- \frac{2 GM}{c^2 r}-\frac{\Lambda r^2}{3}) c^2dt^2 - \frac{1}{(1- \frac{2 GM}{c^2 r}-\frac{\Lambda r^2}{3})} dr^2 -r^2 d\Omega^2,
\end{equation} 
where instead of one, we will have two physical event horizons (EHs)
\begin{equation}\label{EH}
r_1 = \frac{2}{\sqrt \Lambda} cos(\frac{1}{3} cos^{-1}(\frac{3GM \sqrt \Lambda}{c^2})+\frac{\pi}{3}), \quad r_2 = \frac{2}{\sqrt \Lambda} cos(\frac{1}{3} cos^{-1}(\frac{3GM \sqrt \Lambda}{c^2})-\frac{\pi}{3}).
\end{equation}
There is also a third root defined as  $r_3 = -(r_1 + r_2)$ which is not considered as a physical solution. Indeed, the negative solution is regarded as the EH for the spacetime located on the other side of singularity, also known as a ``dual Universe".

\begin{figure}
\centering 
\begin{tikzpicture} 
\node (I)    at ( 2,0)   {I};
\node (II)   at (-2,0)   {};
\node (III)  at (0, 1.25) {II};
\node (IV)   at (0,-1.25) {};
\node (V)   at (4,-1.25) {};
\node (VI)   at (4,1.25) {III};
\node (VII)   at (6,0) {};
\node (VIII)   at (8,1.25) {};
\node (IX)   at (8,-1.25) {};
\node(X) at (10,0) {};

\path  
  (II) +(90:2)  coordinate  (IItop)
       +(-90:2) coordinate (IIbot)
       +(0:2)   coordinate                  (IIright)
       +(180:2) coordinate(IIleft)
       ;
\draw (IIleft) -- 
          node[midway, below, sloped] {$r_2$}
      (IItop) --
          node[midway, below, sloped] {$r_1$}
      (IIright) -- 
          node[midway, above, sloped] {$r_1$}
      (IIbot) --
          node[midway, above, sloped] {$r_2$}
      (IIleft) -- cycle;

\path 
   (I) +(90:2)  coordinate (Itop)
       +(-90:2) coordinate  (Ibot)
       +(180:2) coordinate (Ileft)
       +(0:2)   coordinate (Iright)
       ;
\draw (Ileft) -- 
          node[midway, below, sloped] {$r_1$}
      (Itop) --
          node[midway, below, sloped] {$r_2$}
      (Iright) -- 
          node[midway, above, sloped] {$r_2$}
      (Ibot) --
          node[midway, above, sloped] {$r_1$}
      (Ileft) -- cycle;
\draw  (Ileft) -- (Itop) -- (Iright) -- (Ibot) -- (Ileft) -- cycle;

\path  
  (VII) +(90:2)  coordinate  (VIItop)
       +(-90:2) coordinate (VIIbot)
       +(180:2)   coordinate                  (VIIright)
       +(0:2) coordinate(VIIleft)
       ;
\draw (VIIleft) -- 
          node[midway, below, sloped] {$r_1$}
      (VIItop) --
          node[midway, below, sloped] {$r_2$}
      (VIIright) -- 
          node[midway, above, sloped] {$r_2$}
      (VIIbot) --
          node[midway, above, sloped] {$r_1$}
      (VIIleft) -- cycle;

\path  
  (X) +(90:2)  coordinate  (Xtop)
       +(-90:2) coordinate (Xbot)
       +(180:2)   coordinate                  (Xright)
       +(0:2) coordinate(Xleft)
       ;
\draw (Xleft) -- 
          node[midway, below, sloped] {$r_1$}
      (Xtop) --
          node[midway, below, sloped] {$r_2$}
      (Xright) -- 
          node[midway, above, sloped] {$r_2$}
      (Xbot) --
          node[midway, above, sloped] {$r_1$}
      (Xleft) -- cycle;

\draw[decorate,decoration=zigzag] (IItop) -- (Itop)
      node[midway, above, inner sep=2mm] {$r=0$};

\draw[decorate,decoration=zigzag] (IIbot) -- (Ibot)
      node[midway, below, inner sep=2mm] {$r=0$};

\draw (Ibot) -- (VIIbot)
      node[midway, below, inner sep=2mm] {$r=-\infty$};

\draw (Itop) -- (VIItop)
      node[midway, above, inner sep=2mm] {$r=+\infty$};

\draw[decorate,decoration=zigzag] (VIItop) -- (Xtop)
      node[midway, above, inner sep=2mm] {$r=0$};

\draw[decorate,decoration=zigzag] (VIIbot) -- (Xbot)
      node[midway, below, inner sep=2mm] {$r=0$};

\draw (Xtop) -- (12,2)
	node[midway, above, inner sep=2mm] {$r=+\infty$};
\draw (Xbot) -- (12,-2)
	node[midway, below, inner sep=2mm] {$r=-\infty$};
\draw (IItop) -- (-4,2)
	node[midway, above, inner sep=2mm] {$r=+\infty$};
\draw (IIbot) -- (-4,-2)
	node[midway, below, inner sep=2mm] {$r=-\infty$};

\end{tikzpicture}
\caption{Penrose diagram for SdS BH.} \label{Fig1}
\end{figure}
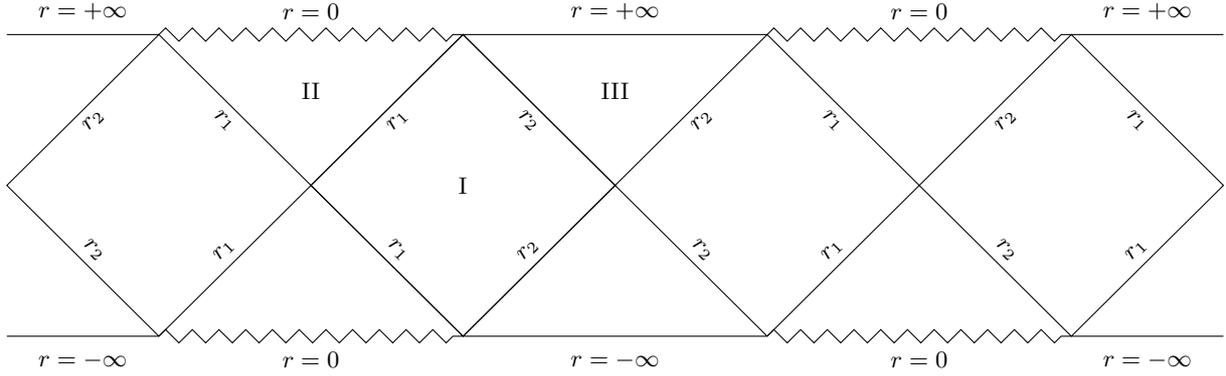

The Penrose diagram of this spacetime is shown in Fig.(\ref{Fig1}).
Here the zigzag lines represent the spacelike singularity inside the BH. In this sense a particle in region \textbf{I}, outside the BH, can either fall into BH i.e. region \textbf{II} or stay outside. Meantime, the region \textbf{III} is the region beyond the cosmological horizon. 

The other possibility for BHs is the Kerr BH which might be formed during the gravitational collapse of stars. In this sense, instead of Eq.(\ref{SdS}), according to ``no-hair theorem" \cite{NH1,NH2,NH3}, one has the following metric

\begin {equation} \label {KerrdS}
ds^2 = \frac {\Delta_r}{\rho^2 L^2}(cdt-a\sin^2\theta d\phi)^2 - \frac{\rho^2}{\Delta_r} dr^2 - \frac{\rho^2}{\Delta_\theta} d\theta^2 - \frac {\Delta_\theta \sin ^2 \theta}{\rho^2 L ^2} (a cdt - (r^2+a^2)d\phi)^2, 
\end {equation}
where
\begin {equation} 
\Delta_r = (1- \frac{\Lambda r^2}{3})(r^2+a^2) - \frac{2 G M r}{c^2}, \quad \Delta_\theta = (1+ \frac{a^2 \Lambda \cos^2 \theta}{3}), \quad L = (1+ \frac{a^2 \Lambda}{3}), \quad \rho^2 = r^2+ a^2 \cos ^2 \theta.
\end {equation}

The above solution is known as Kerr-de Sitter (KdS) metric (see \cite{HG,KS1,KS2,KS3} for detailed analysis of mathematical and physical properties of this metric and its further generalization to charged case). Comparing this metric with Eq.(\ref{SdS}) it turns out that there are two main differences:

-In contrast to SdS metric, the singularity in KdS is timelike and has a ring shape.

-Besides the cosmological horizon $r_C$, we have two other EHs i.e. $r^\pm _{H}$.

\begin{figure}
\centering 
\begin{tikzpicture}
\node (I)    at ( 2,0)   {};
\node (II)   at (-2,0)   {};
\node (III)  at (0, 1.25) {III};
\node (IV)   at (0, -2) {II};
\node (V)   at (4,-1.25) {};
\node (VI)   at (4,1.25) {};
\node (VII)   at (6,0) {};
\node (VIII)   at (8,1.25) {};
\node (IX)   at (8,-1.25) {};
\node(X) at (10,0) {};
\node(XI) at (-2, -4) {};
\node(XII) at (2, -4) {I};
\node(XIII) at (0, -3.25) {};
\node(XIV) at (0, -5.25) {};
\node(XV) at (4, -5.25) {};
\node(XVI) at (4, -3.25) {};
\node(XVII) at (6, -4) {};
\node(XVIII) at (8, -3.25) {};
\node(XIX) at (8, -5.25) {};
\node(XX) at (10, -4) {};
\path  
  (II) +(90:2)  coordinate  (IItop)
       +(-90:2) coordinate (IIbot)
       +(0:2)   coordinate                  (IIright)
       +(180:2) coordinate(IIleft)
       ;
\draw (IIleft) -- 
          node[midway, below, sloped] {$r'_C$}
      (IItop) --
          node[midway, below, sloped] {$r^-_H$}
      (IIright) -- 
          node[midway, above, sloped] {$r^-_H$}
      (IIbot) --
          node[midway, above, sloped] {$r'_C$}
      (IIleft) -- cycle;

\path 
   (I) +(90:2)  coordinate (Itop)
       +(-90:2) coordinate  (Ibot)
       +(180:2) coordinate (Ileft)
       +(0:2)   coordinate (Iright)
       ;
\draw (Ileft) -- 
          node[midway, below, sloped] {$r^-_H$}
      (Itop) --
          node[midway, below, sloped] {$r'_C$}
      (Iright) -- 
          node[midway, above, sloped] {$r'_C$}
      (Ibot) --
          node[midway, above, sloped] {$r^-_H$}
      (Ileft) -- cycle;
\draw  (Ileft) -- (Itop) -- (Iright) -- (Ibot) -- (Ileft) -- cycle;

\path  
  (VII) +(90:2)  coordinate  (VIItop)
       +(-90:2) coordinate (VIIbot)
       +(180:2)   coordinate                  (VIIright)
       +(0:2) coordinate(VIIleft)
       ;
\draw (VIIleft) -- 
          node[midway, below, sloped] {$r^-_H$}
      (VIItop) --
          node[midway, below, sloped] {$r'_C$}
      (VIIright) -- 
          node[midway, above, sloped] {$r'_C$}
      (VIIbot) --
          node[midway, above, sloped] {$r^-_H$}
      (VIIleft) -- cycle;

\path  
  (X) +(90:2)  coordinate  (Xtop)
       +(-90:2) coordinate (Xbot)
       +(180:2)   coordinate                  (Xright)
       +(0:2) coordinate(Xleft)
       ;
\draw (Xleft) -- 
          node[midway, below, sloped] {$r'_C$}
      (Xtop) --
          node[midway, below, sloped] {$r^-_H$}
      (Xright) -- 
          node[midway, above, sloped] {$r^-_H$}
      (Xbot) --
          node[midway, above, sloped] {$r'_C$}
      (Xleft) -- cycle;

\draw[decorate,decoration=zigzag] (IItop) -- (IIbot)
      node[midway, above, inner sep=2mm] {};

\draw[decorate,decoration=zigzag] (Itop) -- (Ibot)
      node[midway, below, inner sep=2mm] {};

\draw[decorate,decoration=zigzag] (VIItop) -- (VIIbot)
      node[midway, above, inner sep=2mm] {};

\draw[decorate,decoration=zigzag] (Xtop) -- (Xbot)
      node[midway, below, inner sep=2mm] {};

\path  
  (XI) +(90:2)  coordinate  (XItop)
       +(-90:2) coordinate (XIbot)
       +(0:2)   coordinate                  (XIright)
       +(180:2) coordinate(XIleft)
       ;
\draw (XIleft) -- 
          node[midway, below, sloped] {$r_C$}
      (XItop) --
          node[midway, below, sloped] {$r^+_H$}
      (XIright) -- 
          node[midway, above, sloped] {$r^+_H$}
      (XIbot) --
          node[midway, above, sloped] {$r_C$}
      (XIleft) -- cycle;

\path 
   (XI) +(90:2)  coordinate (XItop)
       +(-90:2) coordinate  (XIbot)
       +(180:2) coordinate (XIleft)
       +(0:2)   coordinate (XIright)
       ;

\draw  (XIleft) -- (XItop) -- (XIright) -- (XIbot) -- (XIleft) -- cycle;

\path  
  (XII) +(90:2)  coordinate  (XIItop)
       +(-90:2) coordinate (XIIbot)
       +(0:2)   coordinate                  (XIIright)
       +(180:2) coordinate(XIIleft)
       ;
\draw (XIIleft) -- 
          node[midway, below, sloped] {$r^+_H$}
      (XIItop) --
          node[midway, below, sloped] {$r_C$}
      (XIIright) -- 
          node[midway, above, sloped] {$r_C$}
      (XIIbot) --
          node[midway, above, sloped] {$r^+_H$}
      (XIIleft) -- cycle;

\path  
  (XVII) +(90:2)  coordinate  (XVIItop)
       +(-90:2) coordinate (XVIIbot)
       +(180:2)   coordinate                  (XVIIright)
       +(0:2) coordinate(XVIIleft)
       ;
\draw (XVIIleft) -- 
          node[midway, below, sloped] {$r^+_H$}
      (XVIItop) --
          node[midway, below, sloped] {$r_C$}
      (XVIIright) -- 
          node[midway, above, sloped] {$r_C$}
      (XVIIbot) --
          node[midway, above, sloped] {$r^+_H$}
      (XVIIleft) -- cycle;

\path  
  (XX) +(90:2)  coordinate  (XXtop)
       +(-90:2) coordinate (XXbot)
       +(180:2)   coordinate                  (XXright)
       +(0:2) coordinate(XXleft)
       ;
\draw (XXleft) -- 
          node[midway, below, sloped] {$r_C$}
      (XXtop) --
          node[midway, below, sloped] {$r^+_H$}
      (XXright) -- 
          node[midway, above, sloped] {$r^+_H$}
      (XXbot) --
          node[midway, above, sloped] {$r_C$}
      (XXleft) -- cycle;










\end{tikzpicture}
\caption{Penrose diagram for KdS BH.} \label{Fig2}
\end{figure}
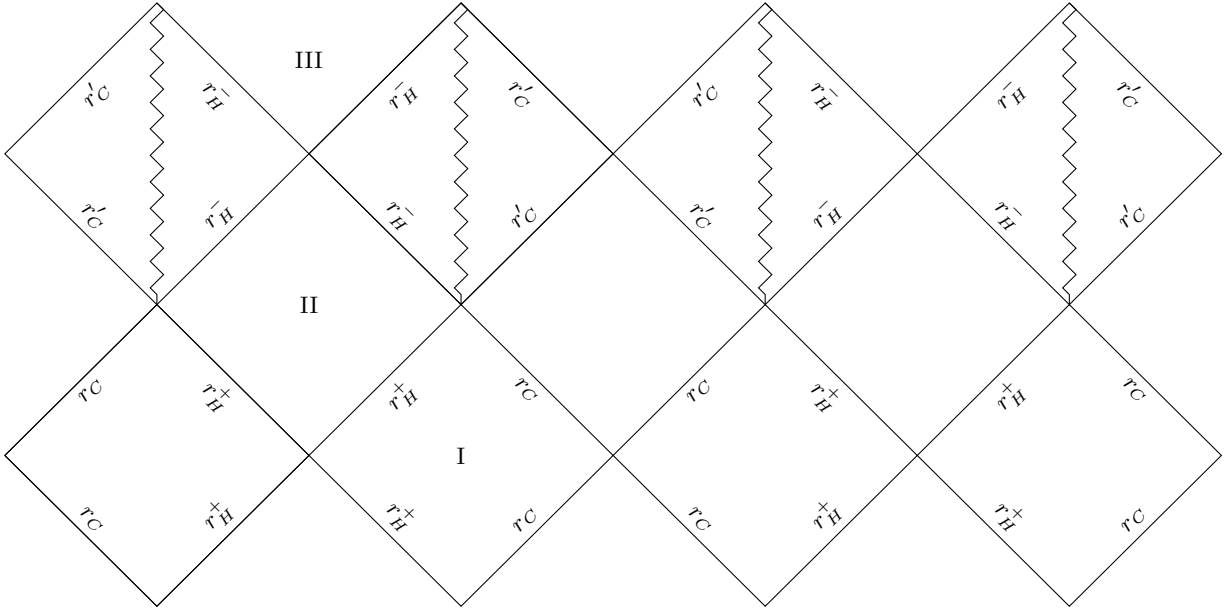

The Penrose diagram of this spacetime is shown in Fig.(\ref{Fig2}).
Here the zigzag lines represent the timelike ring singularity and $r'_C$ denotes the cosmological horizon in $r<0$. Thus, a particle outside the BH, in the region I, can enter into BH, by passing the $r^+_H$, and appear in region II. Then, it will continue its path toward the $r^-_H$. After passing $r^-_{H}$, the other fundamental difference between SdS and KdS becomes evident which we will discuss in the next subsection in more details.

\section{Gedanken experiments with horizons}

Concerning the nature of SdS and KdS solutions, it is possible to have scenarios where several fundamental principles of physics are violated. Namely,

$\bullet$ In case of SdS BHs, the two EHs will become identical if $\frac{3GM \sqrt \Lambda}{c^2}=1$. Surely, considering the numerical values of constants this critical mass is several orders of magnitude larger than supermassive BHs. However, from theoretical point of view there is neither fundamental restriction nor any other prevention to have such an extreme BH. Indeed, according to ``Area Theorem" \cite{BH}, when a particle falls into SdS BH, its information is destroyed in the central (spacelike) singularity. Thus, the area of first horizon $A_1 = 4 \pi r_1^2$ is increased while the area of second one $A_2 = 4 \pi r_2^2$ becomes smaller. Consequently, by continuation of this process finally the two EHs will become identical i.e.
\begin{equation}\label{ext}
r_1 = r_2 = \frac{1}{\sqrt{\Lambda}}.
\end{equation}
In addition, if the accretion to SdS BH continues, finally the extreme SdS BH also will be changed into a full naked singularity.

$\bullet$ In another scenario, consider a BH with mass $M$ which accretes a mass $m_1$. At the same time assume that according to Hawking effect \cite{Hw1,Hw2} an amount of mass $m_2$ is evaporated. In this case, one of the fundamental questions which arises is that 

\textit{What is the output of this mixed process from the BH thermodynamics point of view? Should one treat the process classical or quantum?}.  

However, for both options, we can propose a process which will be contradictory. Namely, by considering the whole process as a classical effect, we can assume a scenario where $m_1 < m_2$. As a result of that we will have a violation of second law of BH thermodynamics i.e. 

\begin{equation}\label{classic}
m_1 < m_2 : \quad S_{initial} = \frac{A_{initial} k_B}{4 l_p^2} > S_{final} = \frac{A_{final} k_B}{4 l_p^2},
\end{equation} 
where $A$ is the area of the first EH. On the other hand, if we consider the whole scenario as a quantum effect, then some internal inconsistencies will happen regarding the First law of BH thermodynamics. Indeed, it should be recalled that the First Law is written as
\begin{equation}\label{1st}
dE = TdS. 
\end{equation}
Accordingly, in this case we will have  

\begin{equation}\label{quantum}
m_1 > m_2 : \quad dE = c^2 d(M + m_1 - m_2) \neq T dS.
\end{equation}
This is due to the fact that while the total amount of increase in both energy/mass on the left hand side and the entropy on the right hand of side Eq.(\ref{quantum}) is related to the amount of $m_1 - m_2$, the radiation is only due to the mass $m_2$ evaporated due to Hawking effect as black body radiation of temperature 
\begin{equation}\label{radiation}
T = \frac{\hbar c^3}{8 \pi G m_2 k_B}.
\end{equation}

$\bullet$
Considering the KdS BH, the situation can be more problematic as there is the risk of violation of ``determinism of GR" which is related to the ``strong cosmic censorship".  Namely, according to Fig.(\ref{Fig2}), since the $r^-_H$ is the Cauchy horizon (CH), it is not possible to determine the evolution of geodesics of infalling particles inside the CH. They can be either destroyed in the singularity or continue their path toward outside (either to region III or $r<0$). Furthermore, it can be easily checked that due to special nature of ring singularity, it is possible that $r^-_H$ starts to increase more rapidly than $r^+_H$. As a result of such process, we will have an extreme KdS with a timelike ring singularity equipped by one EH which is a CH too. The perspective of this scenario will be such that the acausal region (created during the gravitational collapse) i.e. $0<r< r^-_H $ will grow and as a result of that eventually our universe will become completely acausal. Here it should be noticed that, the problem of CH and acausal regions is not restricted to KdS case. Indeed, it covers the whole family of Kerr, Reissner-Nordström (RN) and Kerr-Newman (KN) BH solutions in $\Lambda \geq 0$ asymptote.

\section{Entropy vs horizons}

The physics of BH with multiple horizons has been studied in several papers \cite{MH1,MH2,MH3}. It has been shown in \cite{TSE}, that considering $\Lambda$ in all BH solutions i.e. SdS, RNdS, KdS and KNdS, the total sum of entropies (all horizons, including the negative root) yields a fixed value
\begin{equation}\label{TSE}
\Sigma_{i} S_i = \frac{6 \pi c^3 k_B}{\hbar G \Lambda}.
\end{equation}
Meantime, it can be easily checked that by considering both positive and negative cosmological horizons i.e. $r = \pm \sqrt{\frac{3}{\Lambda}}$, the above relation will be valid for dS spacetime too. Thus, if we take the $\Sigma_{i} S_i$ as the total entropy of the system (BH in dS asymptote), then the Second law of thermodynamics will tell us that this sum must never be decreased. 

Such definition of entropy has the following advantages: 

$\bullet$ First, for extreme and over-extreme cases (where two or even more horizons coincide), the number of horizons are decreased. Thus for such cases we will have
\begin{equation}\label{TSE<}
\Sigma_{i} S_i < \frac{6 \pi c^3 k_B}{\hbar G \Lambda}.
\end{equation}
In this sense, the possible scenario for formation of all types of extreme BHs including the full acausal universe and the naked singularities is dismissed.

$\bullet$
Second, the astrophysical BHs (originally formed as non-extreme KdS), start to loose their rotational energy via Penrose process \cite{PenP} and finally become SdS BHs. Since for both SdS and KdS cases, Eq.(\ref{TSE}) is valid, no violation is occurred during the conversion of KdS to SdS. 

$\bullet$
Third, at the very late time of our Universe, when all KdS BHs eventually become SdS, the Hawking evaporation starts to remove the spacelike singularity of SdS which again does not violate the second law of thermodynamics since as a final result we will obtain the final entropy of dS universe. Namely, by considering Eq.(\ref{SdS}), during the evaporation process we will have
\begin{equation}
M \to 0: \quad 4\pi r_1^2 \to 0 \quad 4 \pi r_2^2 \to \frac{12 \pi}{\Lambda} \quad 4 \pi r_3^2 \to \frac{12 \pi}{\Lambda},
\end{equation}
which means that
\begin{equation}
M \to 0: \quad S_1 \to 0 \quad S_2 \to  \frac{3 \pi c^3 k_B}{\hbar G \Lambda} \quad S_3 \to \frac{3 \pi c^3 k_B}{\hbar G \Lambda}.
\end{equation}

Thus, when the BH singularity is fully removed and BH is evaporated, we will have very low entropy photons (due to thermal radiation of BH evaporation) while the second EH will become identical to horizon of dS universe. The above results are in complete agreement with the Conformal Cyclic Cosmology (CCC), where at the ending phase of an aeon, low entropy photons are produced which can pass through the crossover and have their trace on the next aeon \cite{C1,C2,C3}. Meantime, it should be noticed that in such cases, the negative root  $r_3$ of SdS solution will become identical to $-\sqrt{\frac{\Lambda}{3}}$. Indeed, this negative root can be considered as the dual of our original dS spacetime with the same amount of IC and entropy according to Eq.(\ref{BBdS}).

It should be noticed that, although in the previous section, we have considered the BH solutions in dS asymptote, we can generalize them to FLRW metric too. Indeed, considering the fact that the dS horizon is the special case of Hubble horizon such that
\begin{equation}\label{Hor}
\rho \to 0: \quad \frac{c}{H} \to \sqrt{\frac{3}{\Lambda}},
\end{equation}
($\rho$ is the contribution of all forms of matter, radiation and spatial curvature except the $\Lambda$ term) it is possible to look at the BH solutions in different Cauchy hypersurfaces of FLRW metric. Namely, for each step of IE, we can replace the notion of dS horizon $\sqrt{\frac{3}{\Lambda}}$, by Hubble horizon. Consequently, for BH solutions, once the $j$-th unit of information is created according to Eq.(\ref{BBI}), we will have
\begin{equation}\label{TSEH}
\Sigma_{i} S_i = \frac{2 \pi c^5 k_B}{H_j^2 \hbar G}, \quad H_j^2 = \frac{8 \pi G \rho_{j_{tot}}}{3}.
\end{equation}
Here the $\rho_{j_{tot}}$ is the total amount of density at the $j$-th step. In this sense, at each step one should replace the notion of dS asymptote as the background of BH solution by the background of FLRW which has a constant density equal to $\rho_j = \frac{3c^5}{8j \hbar G^2}$; this treatment can be linked to the Hubble diagram to describe the dark energy models \cite{KS}.

The important consequence of Eq.(\ref{TSEH}) is that, it gives us a {\it universal} notion of entropy, i.e. for all observers in all possible non-extreme BH solutions, for all epochs of universe. Namely, if the observer is located outside the BH,  he/she will consider the total amount of entropy as the $S= \frac{2 \pi c^5 k_B}{H_j^2 \hbar G}$, which is in full agreement with Eq.(\ref{BBI}) (Additional factor 2 is appeared since we have also considered the negative horizon i.e. $r_j= - \frac{c}{H_j}$.). On the other hand, if he/she is inside the BH i.e. between EH  and CH, in the acausal region or even in $r'_C < r <0$, then according to Eq.(\ref{TSEH}) again the same value will be reported.

\section{Bekenstein bound}

Finally, it is worth to mention that originally the Bekenstein bound, Eq.(\ref{BBdS}), was proposed for asymptotically flat spacetimes \cite{BekB2}. However, considering the Hubble horizon and FLRW equations, we can generalize it for curved spacetime too by revisiting the notion of mass and radius. Indeed, this point is essential if we try to check the validity of Eq.(\ref{TSE}) based on Eq.(\ref{BBdS}). Namely for all BH solutions during the IE, if we consider the mass and the EHs of BH i.e.
\begin{equation}\label{MBB}
\Sigma_{i} I_i = \frac{2 \pi M c}{\hbar ln2} (\Sigma_{i} r_i),
\end{equation}  
since we have $\Sigma_{i} r_i = 0$, the above relation will be equal to zero. But, if we consider the contribution of the background, at the moment of creation of  $j$-th unit of information, then
\begin{equation}\label{MBBL}
I=\Sigma_{i} I_i + \Sigma_{j} I_j = \frac{2 \pi M c}{\hbar ln2} (\Sigma_{i} r_i) + (\frac{2\pi}{\hbar ln2})2(\frac{ \pi c^5}{2 j \hbar G^2 })c(\frac{c}{H_j})^4,
\end{equation}
which will be in agreement with Eq.(\ref{TSEH}). It should be noticed that for such cases i.e. BHs in different eras of our Universe, we are dealing with two different objects. First, there are BHs which are characterized by their mass $M$ and their corresponding notions of horizons i.e. $r_{i}$. Second, there is the notion of background with the mass $\pm \frac{ \pi c^8}{2 j \hbar G^2 H^3_j }$ and the characteristic radius i.e. $\pm \frac{c}{H_j}$. (For both mass and radius cases the negative sign represents the corresponding values of dual universe.) Considering both of them in Eq.(\ref{BBdS}), we will obtain Eq.(\ref{MBBL}) where the first term is the contribution of BH in the total IC, while the second term stands for IC of background. In this sense, the universality of entropy for all observers at each epoch in Eq.(\ref{TSEH}) which is based on the summation of areas of all horizons, can be obtained/verified according to Eq.(\ref{MBB}), based on characteristic mass and radii.

\section{Conclusions}

Considering the cosmological constant as a physical constant and the emergence of the information scaling, we studied the entropy of the black holes with multiple event horizons. Using gedanken experiments regarding the Schwarzschild-de Sitter and Kerr-de Sitter black hole horizons, we showed that the information treatment forbids the existence of naked singularities. 

Then we showed that the suggested information approach leads to a {\it universal} notion of entropy, i.e. for all observers and in all possible non-extreme BH solutions and to its link to the Bekenstein bound. The cosmological evolution of the Universe then is represented via a sequence of increasing information defined by means of the physical constants, including the cosmological constant.

\end{document}